\documentclass{aa}

\usepackage{graphicx}
\usepackage{dcolumn}
\usepackage{bm}
\usepackage{txfonts}
\usepackage{natbib}
\usepackage{amsmath}    
\usepackage{graphicx}   
\usepackage{verbatim}   
\usepackage{xcolor}      
\usepackage{subfigure}  
\usepackage{gensymb} 
\usepackage{wasysym} 
\bibpunct{(}{)}{;}{a}{}{,} 
\usepackage{graphicx}
\usepackage{nth}
\usepackage{hyperref}   
\newcommand{\kms}{\mbox{${\rm km\,s}^{-1}$}}

\newcommand{\Msolar}{\mbox{${M}_{\astrosun}$}}
\newcommand{\Rsolar}{\mbox{${R}_{\astrosun}$}}

\newcommand{\Mjup}{\mbox{${M}_{J}$}}

\newcommand{\rhosun}{\mbox{$\rho_{\astrosun}$}}
\newcommand{\Rjup}{\mbox{${R}_{J}$}}
\newcommand{\rhojup}{\mbox{$\rho_{J}$}}

\newcommand\T{\rule{0pt}{2.2ex}}

\begin{document}

\title{Hot Exoplanet Atmospheres Resolved 
with Transit Spectroscopy (HEARTS)\thanks{Based on observations made at ESO 3.6~m telescope (La Silla, Chile) under ESO programmes 098.C-0306 and 100.C-0750 (PI Ehrenreich).}}
\subtitle{VI. Non-detection of sodium with HARPS on the bloated super-Neptune WASP-127b}
\author{J.~V.~Seidel\inst{1} 
\and M.~Lendl\inst{1}
\and V.~Bourrier\inst{1}
\and D.~Ehrenreich\inst{1}
\and R.~Allart\inst{1}
\and S.~G.~Sousa\inst{2}
\and H.~M.~Cegla\inst{1,3}
\and X.~Bonfils \inst{4} 
\and U.~Conod \inst{5} 
\and A.~Grandjean \inst{4}  
\and A.~Wyttenbach\inst{4}
\and N.~Astudillo-Defru\inst{6} 
\and D.~Bayliss\inst{3} 
\and Kevin~Heng\inst{7}
\and B.~Lavie\inst{1}
\and C.~Lovis\inst{1}
\and C.~Melo\inst{8} 
\and F.~Pepe\inst{1}
\and D.~S\'egransan \inst{1}
\and S.~Udry\inst{1}
}
\institute{Observatoire astronomique de l'Universit\'e de Gen\`eve, chemin des Maillettes 51, 1290 Versoix, Switzerland
\and Instituto de Astrof\'isica e Ci\^encias do Espa\c co, Universidade do
Porto, CAUP, Rua das Estrelas, 4150-762 Porto, Portugal
\and Department of Physics, University of Warwick, CV4 7AL Coventry, UK
\and Universit\'e Grenoble Alpes, CNRS, IPAG, 38000 Grenoble, France
\and Department of Physics and Astronomy, University of British Columbia, 6224 Agricultural Road, Vancouver, BC, Canada
\and Departamento de Matem\'atica y F\'isica Aplicadas, Universidad Cat\'olica de la Sant\'isima Concepci\'on, Alonso de Rivera 2850, Concepci\'on, Chile
\and University of Bern, Center for Space and Habitability, Gesellschaftsstrasse 6, CH-3012, Bern, Switzerland
\and European Southern Observatory, Alonso de C\'ordova 3107, Vitacura, Regi\'on Metropolitana, Chile
}

\abstract{WASP-127b is one of the puffiest exoplanets found to date, with a mass of only $3.4$ Neptune masses, but a radius larger than that of Jupiter. It is located at the border of the Neptune desert, which describes the lack of highly irradiated Neptune-sized planets, and which remains poorly understood. Its large scale height and bright host star make the transiting WASP-127b a valuable target to characterise in transmission spectroscopy.
We used combined EulerCam and TESS light curves to recalculate the system parameters. Additionally, we present an in-depth search for sodium in four transit observations of WASP-127b, obtained as part of the Hot Exoplanet Atmosphere Resolved with Transit Spectroscopy (HEARTS) survey with the High Accuracy Radial velocity Planet Searcher (HARPS) spectrograph.
Two nights from this dataset were analysed independently by another team. The team claimed a detection of sodium that is incompatible with previous studies of data from both ground and space. We show that this strong sodium detection is due to contamination from telluric sodium emissions and the low signal-to-noise ratio in the core of the deep stellar sodium lines. When these effects are properly accounted for, the previous sodium signal is reduced to an absorption of $0.46\pm0.20\%$ ($2.3\sigma$), which is compatible with analyses of WASP-127b transits carried out with other instruments. We can fit a Gaussian to the D2 line, but the D1 line was not detected. This indicates an unusual line ratio if sodium exists in the atmosphere. Follow-up of WASP-127 at high resolution and with high sensitivity is required to firmly establish the presence of sodium and analyse its line shape.}
\maketitle

\section{Introduction}

 \begin{table*}
\caption{Observation log.}
\label{table:nightoverview}
\centering
\begin{tabular}{c c c c c c c }
\hline
\hline
&Date   &Retained Spectra \tablefootmark{a}  &Exp. Time [s]  &Airmass        \tablefootmark{b}&Seeing        &S/N order 56      \\
\hline
Night one&2017 February 28&23 (7/16)\tablefootmark{c} &500&1.4-1.1-1.4& - &40 -  50\\
Night two&2017 March 20&33 (16/16)\tablefootmark{d}&600,500&1.9-1.1-1.5&0.7-1.1&35 -  48\\
Night three&2018 February 13&15 (8/6)\tablefootmark{e}&900&1.2-1.1-1.8&0.8-1.2&35 - 55\\
Night four&2018 March 31&33 (13/18)\tablefootmark{f}&600&1.5-1.1-2.7&0.5-1.0&42 - 56\\
\hline
\end{tabular}
\tablefoot{\tablefoottext{a}{In parentheses: spectra in and out of transit, respectively.}
\tablefoottext{b}{Airmass at the beginning, centre, and end of transit.}\tablefoottext{c}{Observed: 37 exposures. Partial transit.}\tablefoottext{d}{Observed: 45 exposures.}\tablefoottext{e}{Observed: 24 exposures. Partial transit, rejected from the analysis for insufficient out-of-transit baseline.}\tablefoottext{f}{Observed: 42 exposures.}}
\end{table*}
Our knowledge of the exoplanet population has increased significantly over the last few years, with thousands of detected exoplanets for a wide range of masses, sizes, and system architectures, some of which are not found in our Solar System. Despite these advances and the consensus that exoplanets are ubiquitous in our Galaxy \citep{Cassan2012,Dressing2013,Wright2012,Howard2013,Batalha2013}, the study of atmospheric composition and structure of exoplanets is still in its early days.

\noindent One of the more important features of the currently known exoplanet population is the so-called Neptune desert, a dearth of highly irradiated planets in the size range of Neptune \citep{Lecavelier2007,Mazeh2016}. This feature is not due to observational bias; these planets should be easily accessible from both space and the ground. A proposed explanation of the Neptune desert is that these planets do not survive the photo-evaporation that is driven by their host stars \citep{Owen2018,Owen2019, Ehrenreich2015, Bourrier2018}. 

\begin{figure}
 \centering
 \label{fig:massinso}
\resizebox{\columnwidth}{!}{\includegraphics[trim=2.5cm 9.0cm 2.5cm 9.0cm]{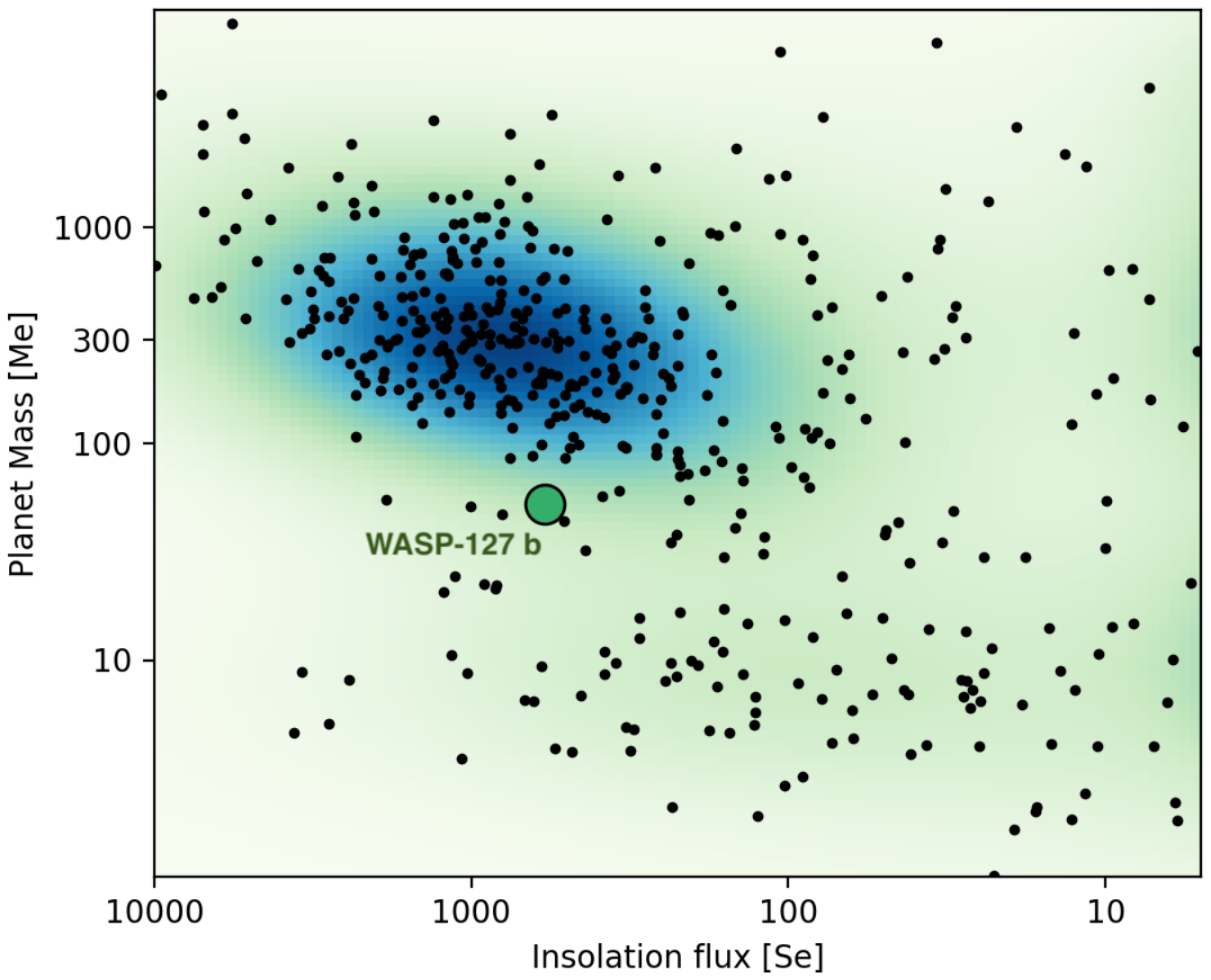}}
 \caption{Exoplanets in the mass vs insolation space. The puffy super-Neptune WASP-127b shown as a big green dot. The data was retrieved from the NASA Exoplanet Archive: \url{https://exoplanetarchive.ipac.caltech.edu/}, retrieved 27 June 2020.}
\end{figure}

\noindent The radius of one of the planets in the Neptune desert, WASP-127b \citep[discovered by the WASP-South survey, ][]{Lam2017}, is comparable to that of Jupiter, but it has a super-Neptune mass, making it one of the puffiest planets to date. WASP-127b is shown in the mass-insolation space in Figure \ref{fig:massinso}. Additionally, its comparatively large radius for a planet of a similar mass cannot be explained by the standard coreless model \citep[see][]{Fortney2007}.

\noindent WASP-127b orbits its bright G5-type host star (V=10.2) with a period of 4.18 days, but it receives a relatively low XUV flux despite its close-in position in the system \citep{Chen2018}. This makes photo-evaporation an unlikely explanation for its inflation. Potential alternatives include re-inflation by the host star, enhanced atmospheric opacity, and Ohmic or tidal heating \citep{Perna2010,Leconte2010, Batygin2010, Batygin2011, Rauscher2013, Lithwick2014, Lammer2016}. At the time of writing, none of these concepts can be established as the clear explanation for the puffiness of WASP-127b, and more precise observations of its atmospheric composition \citep[e.g.][]{dosSantos2020} and system parameters are needed.

Fortunately, WASP-127b is a prime candidate for spectroscopic observations, with a bright host star and an enormous atmospheric scale height \citep[$\sim 2350$ km, ][]{Lam2017}. Previous observations with the 2.5m Nordic Optical Telescope (NOT) and the Gran Telescopio Canarias (GTC), see \cite{Palle2017} and \cite{Chen2018}, have shown evidence for sodium, lithium, potassium, hazes in the form of a Rayleigh slope, and tentative indications for TiO and VO. Space-based observations using the Hubble Space Telescope (HST) and the Spitzer telescope conducted by \cite{Spake2019} were able to detect $\rm H_2O$ and $\rm CO_2$ bands. The observations restricted the metallicity of WASP-127b to super-solar, which has recently been confirmed, together with the cloudy nature of the atmosphere by an independent study using HST data \citep{Skaf2020}. Additionally, the results reported by \cite{Spake2019} are in line with the ground-based sodium observations. 

\noindent  \cite{Zak2019}, from here Z19, reported a sodium signal with an absorption of more than $1\%$ in two transit observations of HARPS (High Accuracy Radial velocity Planet Searcher) data, which were observed as part of the HEARTS survey. The two transit observations comprised a total of four transits of WASP-127b. The analysis of the transits corresponds to a sodium detection of $0.1060\pm0.0008\,R_p/R_\star$ on the wavelength grid from \cite{Spake2019}, which differs from the \cite{Spake2019} results at the $1\,\sigma$ level ($\sim 0.1035\pm0.0010\,R_p/R_\star$, see Figure 21 in \cite{Spake2019}). In the following, we discuss the full available HARPS dataset for WASP-127b of four observation nights and show the effect of telluric sodium and low signal-to-noise ratio (S/N) traces stemming from the stellar sodium lines when planets with low transit depths are observed. Additionally, we present updated system parameters obtained from photometric observations with EulerCam at the Swiss telescope in la Silla, Chile, and the Transiting Exoplanet Survey Satellite (TESS).

\section{Observations and data reduction}
\label{sec:obs}
As part of the HEARTS survey (ESO programme: 098.C-0306, 100.C-0750; PI: Ehrenreich), we observed four transits of the inflated super-Neptune WASP-127b in front of its host star. The transits took place on 2017 February 28, 2017 March 20, 2018 February 13, and 2018 March 31 and were observed with the HARPS spectrograph at the ESO 3.6 m telescope in La Silla Observatory, Chile \citep{Ma03}.

\noindent Because of visibility constraints of the target, two of the transits were only partially recorded and only have out-of-transit spectra before or after the transit itself. The out-of-transit spectra are defined as spectra that were recorded when no part of the planet was in front of its host star. All spectra in which the planet is partially or completely in front of the star are in-transit spectra. In four nights, a total of 141 spectra were taken, 86 in-transit and 55 out-of-transit. An observation log can be found in Table \ref{table:nightoverview}. For each night, we discarded the exposures in which the low S/N core of the stellar lines overlap the planetary orbital track by too much. This concerns $10,6,5,\text{and }7$ exposures in the four observation nights, respectively (see Section \ref{sec:lowS/N} and Appendix \ref{app:2Dmaps}, as well as \cite{Seidel2020b}, for further details). In night 1 a partial transit was recorded, of which one spectrum at the end (in-transit) was only partially taken and not discarded, but added to the transit. In the same night, three in-transit spectra showed significantly lower S/N, most likely due to passing clouds. These spectra were discarded for this analysis. In night two, seven consecutive in-transit spectra were rejected for low S/N, most likely again due to a passing large cloud. Of night three, in which a partial transit was recorded, two out-of-transit exposures were rejected because they were taken to test the best exposure time and not as part of the transit. Additionally, two in-transit spectra were rejected because their S/N was significantly lower than that of the remaining spectra (unknown origin). This left an insufficient number of out-of-transit spectra to properly correct the stellar spectrum, and night three was subsequently rejected from the analysis. In night four, four in-transit spectra were rejected because their S/N was low.

This work focusses on the sodium doublet, which in the case of HARPS is located in echelle order $56$, spanning the wavelength region from $5850.24$~\r{A} to $5916.17$~\r{A}. Fibre A was set on the target during the transit, and Fibre B observed the sky. The recorded HARPS spectra were extracted from the e2ds raw data by the HARPS Data Reduction Pipeline (DRS v3.5)\footnote{\url{http://www.eso.org/sci/facilities/lasilla/instruments/harps/tools/archive.html}}. The wavelength solution is given in the Solar System barycentric rest frame. Additionally, the data were corrected for the blaze, cosmic particles hitting the detector, and telluric contamination. 

Telluric lines play a major role in transmission spectroscopy. Depending on the velocities of the observed system, the telluric lines can overlap with the stellar sodium doublet, which distorts the line shape and depth. To correct for this effect, we used {\tt molecfit} version 1.5.1. \citep{Sm15, Ka15}, an ESO tool for correcting telluric features in ground-based observations. For further details on how to apply {\tt molecfit}, see \cite{Al17}, and for examples in the wavelength range of the sodium doublet, see \cite{Casasayas2019,Seidel2019,Chen2020, Casasayas2020,Hoeijmakers2020} or \cite{Seidel2020b}. The telluric lines were corrected down to the noise level for all airmasses. However, molecfit only corrects for local atmospheric profiles based on models and the measured weather data at the local site, but not for sky emission. It therefore cannot correct for telluric sodium emission lines. This potential contamination can be estimated from the data taken with Fibre B, which was set to the sky. We discuss the effect of telluric sodium in Section \ref{sec:stellar}.

\section{Simultaneous photometry with EulerCam}
\label{sec:EulerCam}

\begin{table}
\centering   
\caption{\label{tab:phot}Overview of the EulerCam photometric observations.}
\begin{tabular}{llll} \hline \hline
date & filter &  $\mathrm{RMS_{2min}}$ [ppm] & noise added\T \\
\hline
2014 April 28 & r'-Gunn & 0.000959  &  0.000625   \\
2016 March 09 & r'-Gunn & 0.000910  &  0.000812  \\
2017 February 28 & r'-Gunn  & 0.000704  &  0.000733  \\
2017 March 20 & r'-Gunn  & 0.000721  &  0.000590  \\
2018 February 13 & r'-Gunn  & 0.001069 &  0.000953  \\
2018 March 31 & r'-Gunn  & 0.000946  &  0.000769  \\
2019 February 24 & r'-Gunn  & 0.000946  & 0.000709   \\
\hline
\end{tabular}
\end{table}

\begin{figure}
 \centering
\resizebox{\columnwidth}{!}{\includegraphics[trim=0.5cm 0.5cm 0cm 0cm]{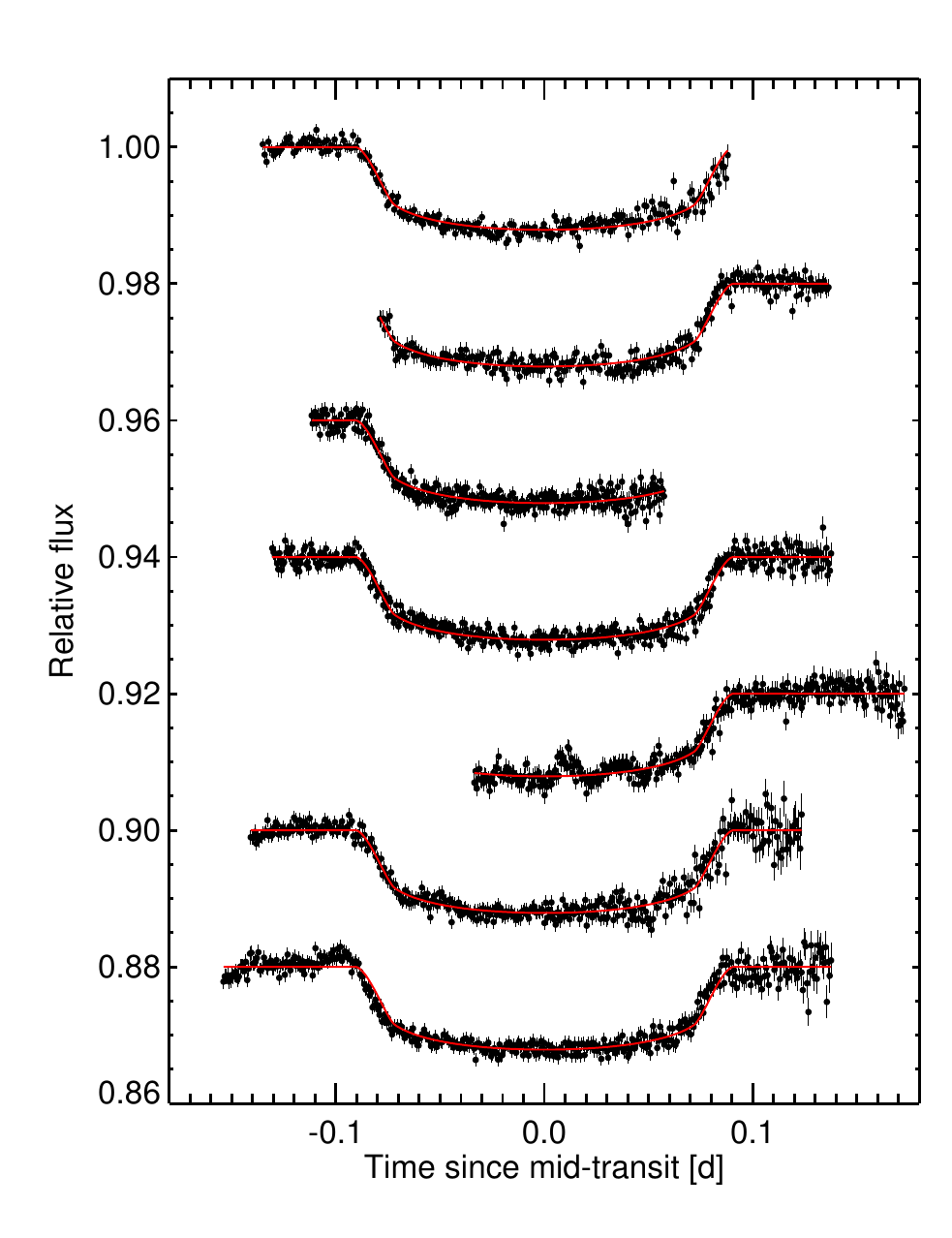}}
 \caption{Light curves obtained by EulerCam (bottom to top: 2014 April 28, 2016 March 09, 2017 February 28, 2017 March 20, 2018 February 13, 2018 March 31, 2019 February 24 offset by 0.02 for visibility. The computed models are shown in red.}
  \label{fig:Ecam}
\end{figure}

\begin{figure}
 \centering
\resizebox{\columnwidth}{!}{\includegraphics[trim=0.5cm 0.5cm 0cm 0cm]{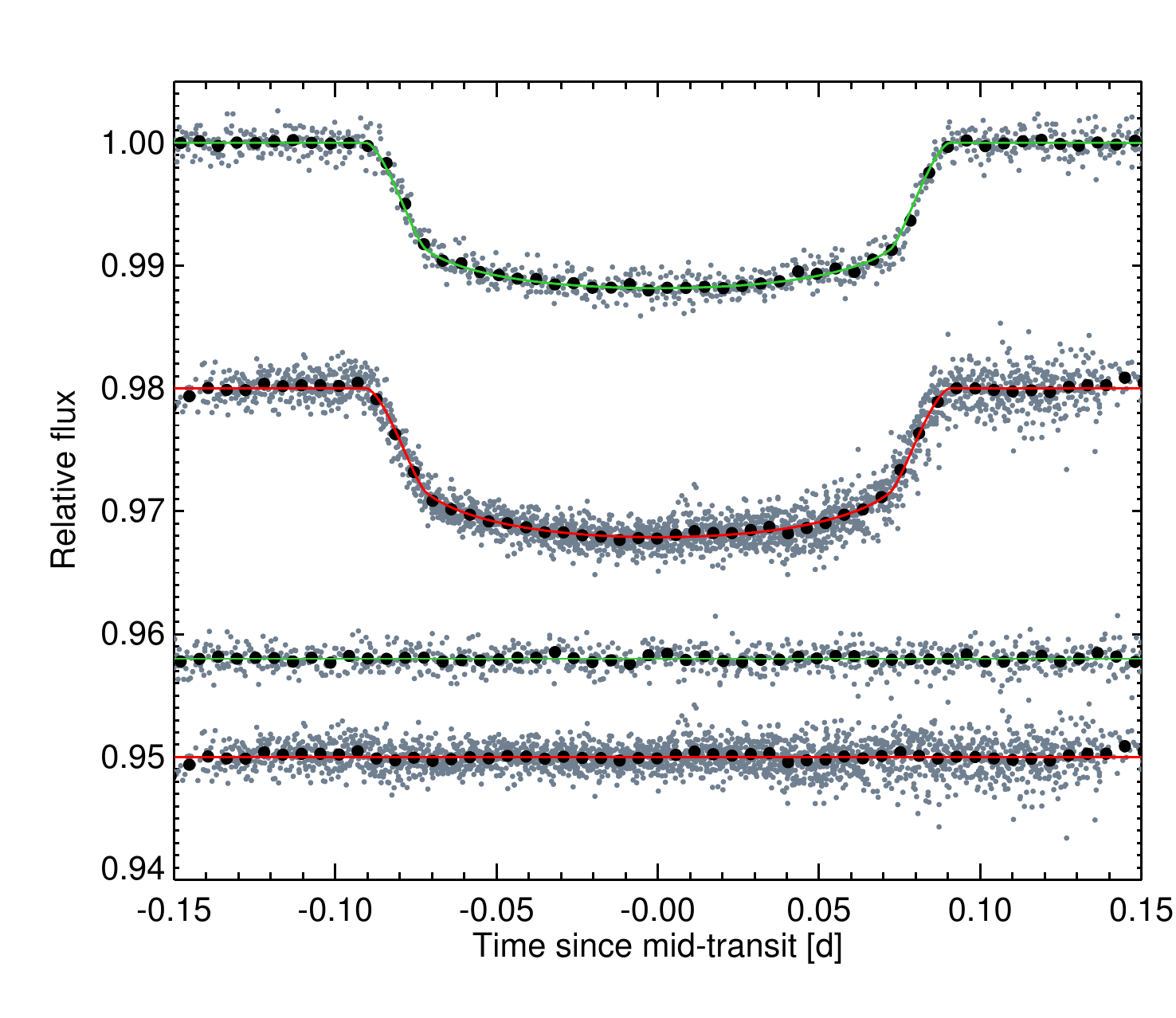}}
 \caption{Light curves obtained by EulerCam (red model fit) and TESS (green model from detrended data and best-fit). The residuals from fitting are shown in the bottom, respectively.}
 \label{fig:photcomp}
\end{figure}

We obtained photometric observations with EulerCam, the charged coupled device (CCD) imager installed at the 1.2 m Euler telescope at La Silla Observatory, Chile. A detailed description of the instrument, as well as the processes needed to obtain relative aperture photometry, can be found in \cite{Lendl2012}. For each transit event, stable reference stars were chosen iteratively. The   light curves of full or partial transits thus obtained are shown in Figure \ref{fig:Ecam}. An overview of the main parameters for each observations can be found in Table \ref{tab:phot}. The photometric uncertainties of the EulerCam data do not reflect additional correlated noise due to stellar, atmospheric, or instrumental effects. To compensate for this, we added the values given in Table \ref{tab:phot} quadratically to the uncertainties of each data set, enforcing a reduced $\chi^2$ of unity for each light curve. The combined 2 min EulerCam RMS is $0.000261$. Additionally, one TESS light curve was observed in Sector 9, covering four individual transits, with a combined 2 min RMS of $0.000209$. The PDC-SAP TESS light curve \citep{Smith2012,Stumpe2014,Jenkins2016} was included in our analysis, and a comparison of combined EulerCam light curves and the TESS light curve is shown in Figure \ref{fig:photcomp}.

\section{Revised system parameters}
\label{sec:paramphot}

We analysed all available photometric transits with the 'code for transiting exoplanet analysis' (CONAN) \citep{Lendl2020}, which is based on the transit analysis code presented in \cite{Lendl2017} and uses the algorithms from \cite{Kreidberg2015} to calculate transit models. The fitted parameters (MCMC jump parameters) are listed in the top section of Table \ref{tab:para} and the model fit is shown as red lines in Figure \ref{fig:photcomp}. 

We used a combined HARPS spectrum to derive the stellar spectroscopic parameters ($T_{\mathrm{eff}}$, $\log g$, microturbulence, [Fe/H]), and respective uncertainties. For this purpose, we followed the ARES+MOOG method as described in \cite{Sousa-14} and \cite{Santos-13}. The ARES code \footnote{The last version of ARES code (ARES v2) can be downloaded at http://www.astro.up.pt/$\sim$sousasag/ares} \citep{Sousa-07, Sousa-15} was used to consistently measure equivalent widths (EW) of iron lines that are included in the line list presented in \citet[][]{Sousa-08}. Briefly, ARES+MOOG performs a minimisation process searching for the ionisation and excitation equilibrium to find convergence on the best set of spectroscopic parameters. To compute the iron abundances, we used a grid of Kurucz model atmospheres \citep{Kurucz-93} and the radiative transfer code MOOG \citep{Sneden-73}. The stellar mass, radius, and age were obtained through the use of the Padova stellar model isochrones \citep[see.][]{daSilva-06, Bressan-12} - http://stev.oapd.inaf.it/cgi-bin/param\_1.3) by providing our spectroscopic parameters together with the Gaia DR2 parallax and the V magnitude. The limb-darkening values were derived using the routines by \cite{Espinoza2015}.
All updated stellar and planetary parameters are given in Table \ref{tab:para} and are used for the spectroscopic analysis in this work.

\begin{table}[h]
\centering   
\caption{\label{tab:para}Planetary and stellar parameters obtained from EulerCam and TESS photometry.}
\begin{tabular}{ll} \hline \hline
\multicolumn{2}{l}{Jump parameters:} \T  \\
\hline
Mid-transit time,[BJD] - 2450000                \T & $  6776.62124_{-0.00028}^{+0.00023}                   $ \\    
 $ R_{p}/R_{\ast} $                             \T & $  0.10103_{-0.00047}^{0.00026}                      $ \\                    
$ a/R_{\ast} $                                   \T & $ 7.81 _{-0.09}^{+0.11}                 $ \\                                                                                    
Period [d]                                      \T & $  4.17806203 _{-0.00000053}^{0.00000088} $ \\                          
RV amplitude, $K\, [\kms]$              \T & $  0.022 _{-0.002}^{0.003} $ \\                                                      
\hline                                             
\multicolumn{2}{l}{Primary parameters from spectral analysis:} \T   \\
\hline                                                                                                          
Stellar radius, $ R_{\ast} $ [{\Rsolar}]           \T & $ 1.335  _{-0.029}^{+0.025}                 $ \\                                                    
Stellar mass, $ M_{\ast} $ [{\Msolar}]           \T & $ 0.949   _{-0.019}^{+0.022}               $ \\                                                      
\hline                                               
\multicolumn{2}{l}{Derived parameters:} \T                                                        \\
\hline                                             
Planetary radius, $ R_{p} $ [{\Rjup}]            \T & $ 1.311  _{-0.029}^{+0.025}               $ \\                                                      
Planetary mass, $ M_{p} $ [{\Mjup}]            \T & $ 0.1647  _{-0.0172}^{+0.0214}               $ \\                                                      
Planetary mean density, $ \rho_{p} $ [{\rhojup}] \T & $ 0.073_{-0.009}^{+0.010}              $ \\                                       
Transit duration [d]                            \T & $  0.18137_{-0.00058}^{+0.00035}        $ \\                                                             
Transit depth, $ \Delta F$ \T                    \T & $0.01021_{-0.00010}^{+0.00005}           $ \\                                          
Impact parameter                                \T & $  0.29 \pm 0.04               $ \\                                                                                                 
$ a/R_{\ast} $                                   \T & $ 7.81 _{-0.09}^{+0.11}                 $ \\                                                                                    
Orbital semi-major axis, $ a $ [au]              \T & $ 0.04840_{-0.00095}^{+0.00136}           $ \\                                  
Stellar mean density, $ \rho_{\ast} $ [{\rhosun}] \T & $ 0.401_{-0.026}^{+0.025}              $ \\                                       
Stellar radius, $ R_{\ast} $ [{\Rsolar}]           \T & $ 1.333 \pm 0.027                $ \\                                                     
Stellar mass, $ M_{\ast} $ [{\Msolar}]           \T & $ 0.950  \pm 0.020               $ \\                                                      
Inclination [deg]                                \T & $ 87.84  _{-0.33}^{+0.36}                 $ \\                                                                            
Eccentricity, $ e $ (fixed)                      \T & $ 0.0                                     $ \\                                              
\hline  
\multicolumn{2}{l}{Fixed quadratic limb-darkening parameters:} \T   \\
\hline  
  $u_{1,\rm r} $      \T & $0.422               $ \\                                                                                                                            
 $ u_{2,\rm r} $      \T & $ 0.214            $ \\
 $ u_{1,\rm t} $     \T & $ 0.350              $ \\
 $ u_{2,\rm t} $     \T & $ 0.225                      $ \\
$ c_{1,\rm r}=2u_{1,\rm r}+u_{2,\rm r} $        \T & $         1.057     $ \\                                    
$ c_{2,\rm r}=u_{1,\rm r}-2u_{2,\rm r} $        \T & $        0.208    $ \\
$ c_{1,\rm t}=2u_{1,\rm t}+u_{2,\rm t} $     \T & $         0.925        $ \\
$ c_{2,\rm t}=u_{1,\rm t} -2u_{2,\rm t} $    \T & $         0.125         $ \\
 \hline
\end{tabular}
\end{table}

\section{Transmission spectroscopy of WASP-127b}
\label{sec:fulltrans}

\label{sec:spec}

The goal of transmission spectroscopy is to extract the planetary signal in the in-transit spectra. The base of this procedure is to sum all in-transit spectra to produce the master-in and divide it by the master-out, thus taking out the stellar spectral lines from the transmission spectrum $\mathfrak{\tilde{R}}$ \citep{Br01}.
However, ground-based observations are not suited for the direct application of this method because the flux changes with time as a result of the varying airmass. Each in-transit spectrum was therefore fitted by a third-order polynomial and then divided by the master-out in the stellar rest frame to guarantee comparable flux levels in all spectra. These corrected in-transit spectra were then shifted to the planet rest frame to align the planetary sodium signal, and they were summed together for higher S/N \citep{Wy15}. The full description of this method can be found in \cite{Seidel2019}.

Two of the four transits (nights one and two) from our programme have been analysed in Z19. Z19 analysed the two transits taken in 2017, which were at the time publicly available in the archive, corresponding to 82 spectra (52 in-transit). Based on this subset of our data, Z19 claimed a sodium detection at the $8.33\sigma$ (D2) and $4.2\sigma$ level (D1) (see Figure 5 in Z19)\footnote{In Z19, the D1 and D2 lines have switched labels, which we assume is a typo.}. In the following we reproduce this analysis and highlight how telluric sodium emissions and low S/N remnants can produce false-positives. 

\subsection{Contamination sources in the sodium wavelength range}
\label{sec:stellar}

\begin{figure}[t]
\resizebox{\columnwidth}{!}{\includegraphics[trim=3.0cm 9.2cm 3.2cm 9.5cm]{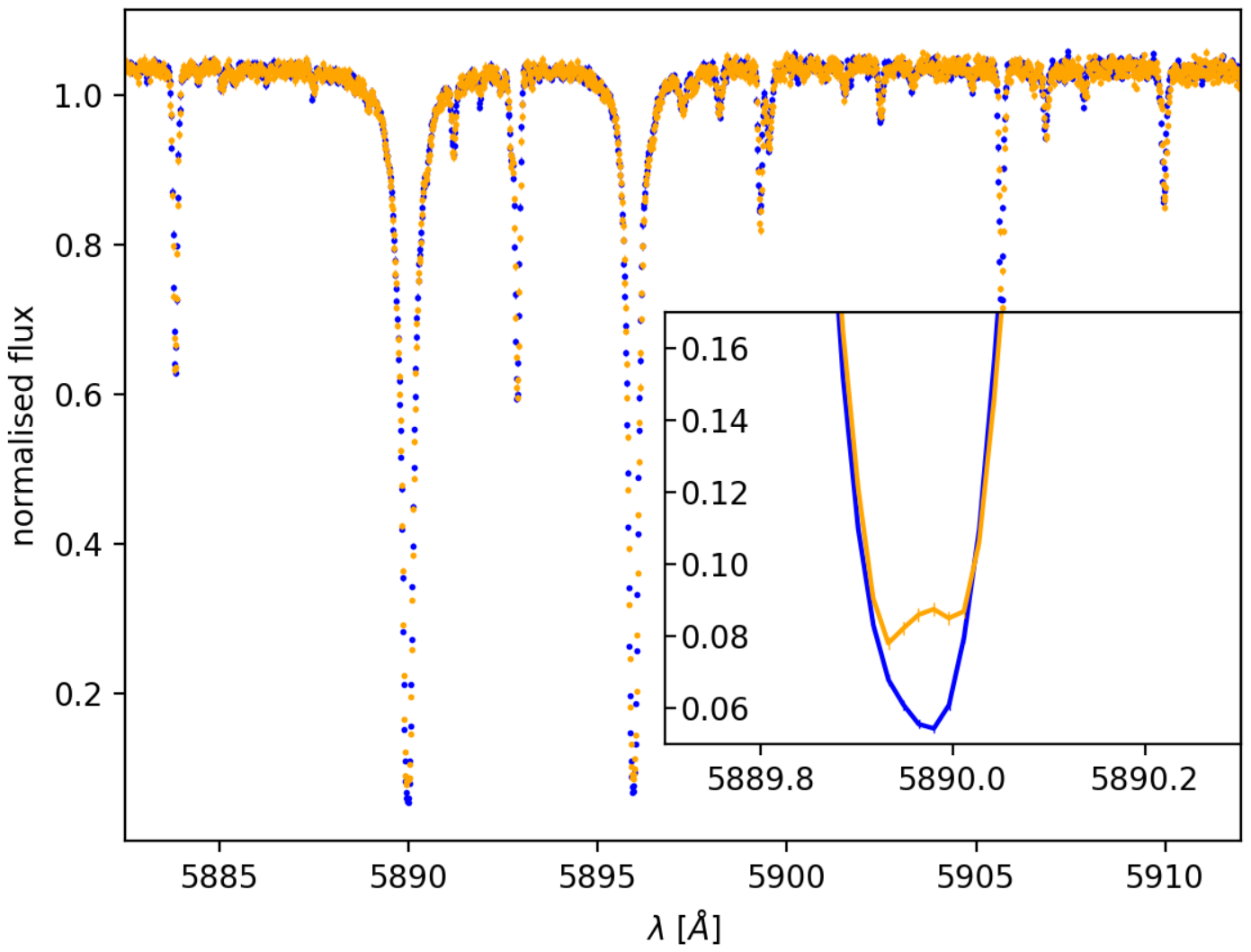}}
        \caption{Normalised flux of the master out in the wavelength range of the sodium doublet. The stellar sodium doublet of night one is shown in blue, of night two in orange. The inset shows a zoom into the line centre of the D2 line of the same data set. Night one shows the expected line shape of the stellar sodium doublet, however, night two shows a sodium emission at line centre (width approx. 6-8 bins), which has to be masked in each spectrum to avoid a false-positive detection of the stellar sodium feature as planetary sodium.}
        \label{fig:masterout_stellar}
\end{figure}

When we look at the data of each night individually, the sodium detection is curiously strong in night two, while a tentative sodium detection is seen in night one, but not in night four. Because at the time of analysis, Z19 only had access to the first two nights and because night one is a partial transit, their sodium detection is dominated by data from night two. An in-depth analysis of night two reveals an emission feature in the master-out sodium line core when compared with the master-out from the other three nights (see Figure \ref{fig:masterout_stellar}).  This emission causes the stellar line to appear more shallow. When the too shallow stellar sodium line is then used to extract the planetary sodium by division of the master-out, it creates an artificially deep residual because the emission varies in time and is not necessarily the same in and out of transit. This explains the unusually deep sodium feature in night two. This emission feature stems from telluric sodium, which was detected on fiber B in nights two and four. 

To correct for the effect of the telluric sodium emission, we masked out the wavelength range of the emission feature in the stellar rest frame in each spectrum of nights two and four. The masking window was set at $\pm3.5\,\kms$ of each line centre, based on the bin size of HARPS and the number of bins affected by the telluric sodium emission, see Figure \ref{fig:masterout_stellar}. For WASP-127b, the planet signal in the stellar rest frame is shifted between $-40\,\kms$ and $20\,\kms$, with changes larger than one $\kms$ from one exposure to the next. This implies that the masking will crop out part of the planetary sodium signal when it overlaps the masked region, in this case, in three to four exposures around mid-transit. However, we discuss Section \ref{sec:lowS/N} why spectra around mid-transit have to be rejected: they are affected by low S/N remnants. As a result, the masking for telluric sodium emission does not reduce the signal strength.

\begin{figure}[t]
\resizebox{\columnwidth}{!}{\includegraphics[trim=3.0cm 9.2cm 3.2cm 9.5cm]{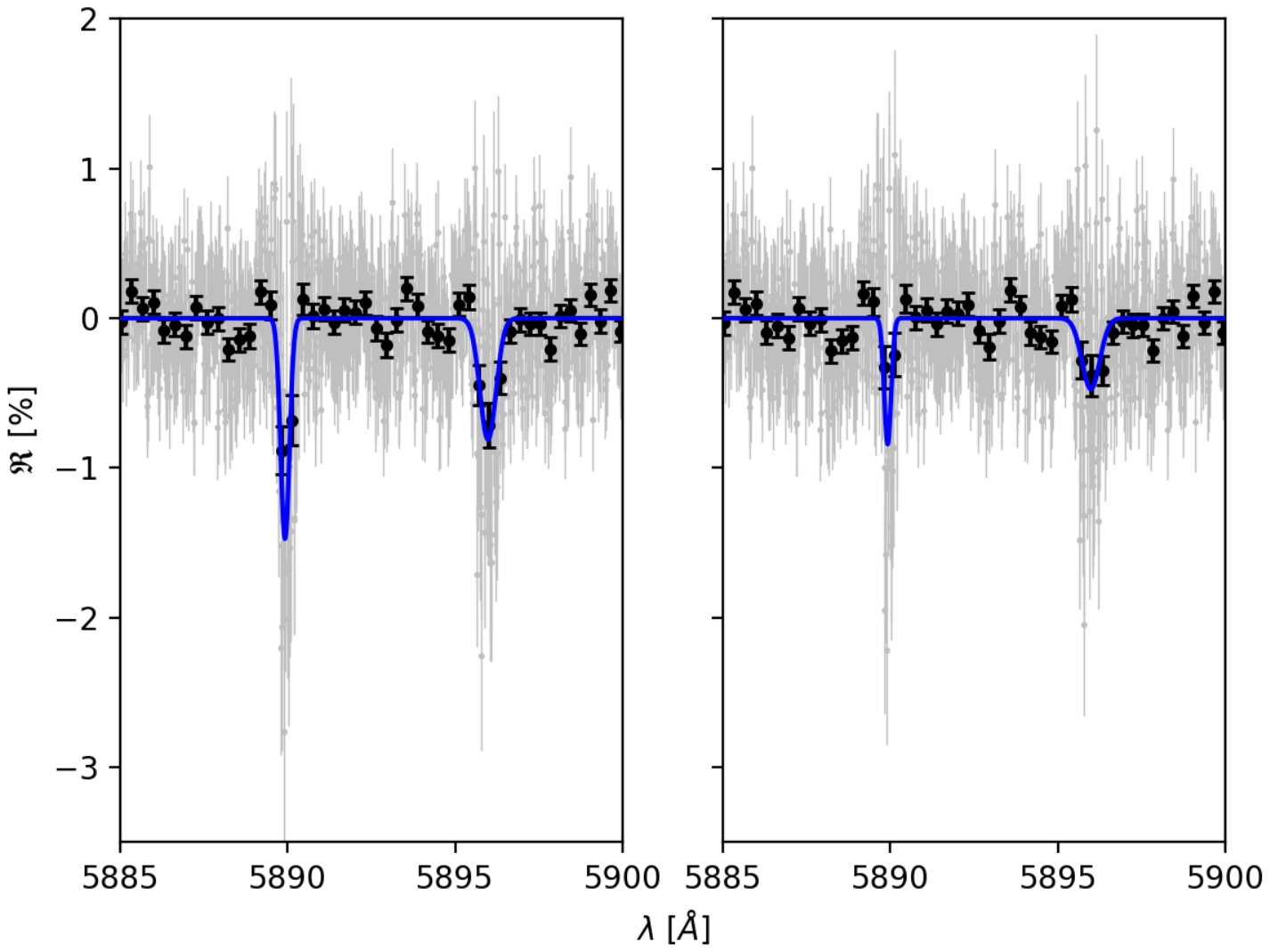}}
        \caption{Percentage of extra absorption as shown in Z19, Figure 5 for the wavelength area of the sodium doublet. Grey points are at HARPS resolution, black points binned by 20. In the left panel is the transmission spectrum of WASP-127b created from the first two transits, that were available to the authors of Z19. On the right, the same dataset is taken and the core of the stellar sodium doublet masked out (we masked $\pm3.5\,\kms$ of each line centre, amounting to roughly 8-9 bins per spectrum). In blue is a Gaussian fit on the grey data for visualisation. In line with Z19, we did not apply any spectra selection. The lower noise in comparison with Z19 can be attributed to our automatic correction for cosmics and different telluric corrections.}
        \label{fig:Zak_stellar}
\end{figure}

In Figure \ref{fig:Zak_stellar} we recreate the sodium transmission spectrum of Z19 (created from nights one and 2) at the left, but with our own telluric correction, and show the effect of masking the region that is affected by telluric sodium emission on their analysis at the right, with a significant decrease in the sodium feature. In total, the telluric sodium contamination makes up $43\%$ of the signal strength in Z19, which reduces the detection levels from 8.33 to 4.8$\sigma$ for the D1 and from 4.2 to 2.4$\sigma$ for the D2 line. Out of all four nights, nights two and four were affected by telluric sodium emission and subsequently corrected in our analysis. 

A stellar effect that might affect the transmission spectrum is the Rossiter-McLaughlin (RM) effect \citep{ Rossiter1924, McLaughlin1924, Queloz2000, Lo15, Cegla2016}. This effect describes the impact of the Doppler-shift from the radial velocity of the local region of the stellar surface occulted by the planet at a given time of its transit. This velocity is below $\sim 2 \kms$ for WASP-127. \cite{Wy17} showed that the RM effect can be neglected for very slowly rotating stars. We thus concur with the reasoning of Z19 to neglect the RM effect in the analysis of the dataset presented here.

\begin{figure*}[htb]
\resizebox{\textwidth}{!}{\includegraphics[trim=0.5cm 9.0cm 0.5cm 9.2cm]{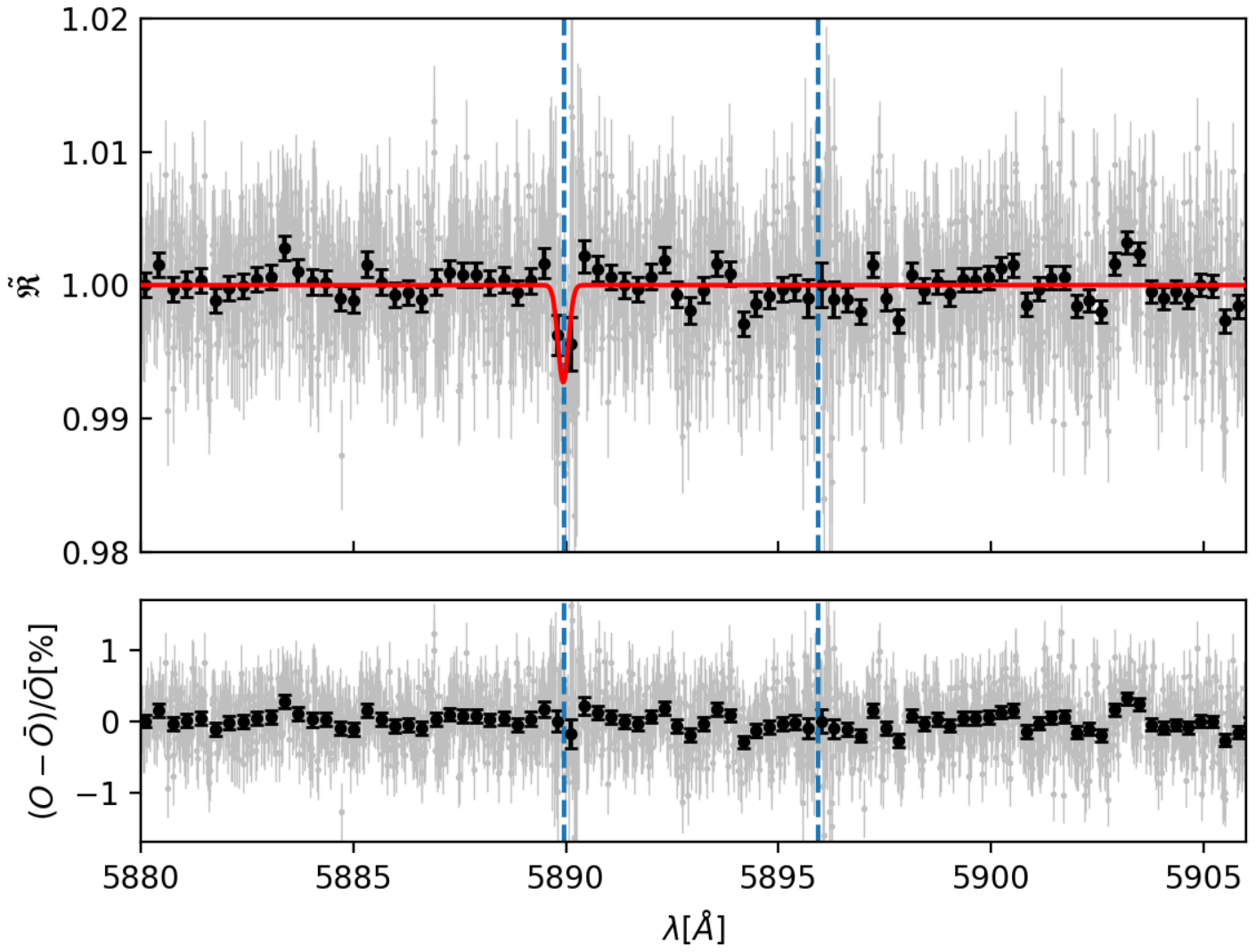}}
        \caption{HARPS transmission spectrum of WASP-127b for the sodium doublet in the planetary rest frame. Upper panel: The transmission spectrum for all three (1,2,4) nights combined in full HARPS resolution is shown in grey, in black the same data is shown binned by x20. The data has been corrected for tellurics, cosmics, telluric sodium emission contamination and low S/N remnants. The rest frame transition wavelengths are marked with blue dashed lines. We attempted a Gaussian fit to both potential sodium lines jointly, however, the fit was only successful for the D2 line. The fit is shown as a red line. We, therefore, fitted the continuum to the D1 line instead. Lower panel: Residuals of the Gaussian fit. }
        \label{fig:transspectrum}
\end{figure*}

\subsubsection{Stellar variability tracers}

In addition to the telluric and stellar effects described before, stellar activity can affect the sodium feature. To detect any stellar effects, the Ca I and Mg I lines were monitored in Z19 (see Figure 3 in Z19), and no signal was found. Based on these results, Z19 assumed no effect of stellar activity in their analysis. 
However, we remark that these measurements have a low S/N, likely due to the inclusion of spectra that were taken while clouds passed or that were discarded for testing purposes (see Table 2 in Z19). Additionally, the plot of the Mg I line in Figure 3 of Z19 shows a spread of the data points at line centre greater than the noise level, with an uncertainty about twice as large at line centre than in the continuum. The high noise level in the Mg I line and the overall noise could effectively mask a signal and hide stellar activity. When we assume that no signal was hidden for Ca I and Mg I during the combined analysis of night one and 2, temporal stellar activity might still affect the transmission spectrum, but not be visible in the time-integrated spectrum. 
In this work, we present EulerCam light curves taken simultaneously with the spectroscopic data, see Figure \ref{fig:Ecam}, which show no photometric variability that might be attributable to stellar activity (e.g., spot modulations).

\subsubsection{Low S/N trace}
\label{sec:lowS/N}
In each stellar spectrum, the most dominant feature is stellar sodium, where the flux drops significantly in the line core (see Figure \ref{fig:masterout_stellar} for an example). Because of the low flux, the photon noise is increased compared to the signal, resulting in a low S/N ratio in the wavelength region of the stellar sodium doublet. The planetary signal is extracted from the total flux by dividing each spectrum by the master-out in the stellar rest frame (SRF). In consequence, low S/N residuals remain at the former position of the stellar sodium doublet \citep{Barnes2016, Borsa2018,Seidel2020b}. When the spectra are then shifted to the planetary rest frame (PRF), the low S/N residuals are also shifted to different positions for each exposure (see Appendix \ref{app:2Dmaps} for 2D maps of the transits). However, for part of the in-transit spectra, the low S/N residuals lie in the wavelength range of the potential planetary sodium. In these regions, the residuals can mask a sodium signal due to the increase in noise or even produce a false sodium signature.
We mitigate this effect by identifying the overlap between the low-S/N core of the stellar line (the FWHM of the stellar sodium feature at the wavelength sodium is known to absorb light) and the spectral range that potentially contains the planet sodium signal in each exposure. For WASP-127b this corresponds to a phase range from $-0.008$ to $0.008$ in which this overlap would dominate any potential sodium signal. We subsequently discarded the affected exposures, see Table \ref{table:nightoverview}. A closer look at the 2D maps in Appendix \ref{app:2Dmaps} shows that the trace in night one and especially in night two is mainly in absorption and could thus introduce a false signal. This would account for the additional difference between the analysis in Z19 (barring the telluric sodium contamination) and our work. Night four is largely unaffected by the low S/N trace because the S/N is higher overall.

\subsection{Sodium in WASP-127b}

\begin{figure*}[htb]
\resizebox{\textwidth}{!}{\includegraphics[trim=0.5cm 9.0cm 0.5cm 9.1cm]{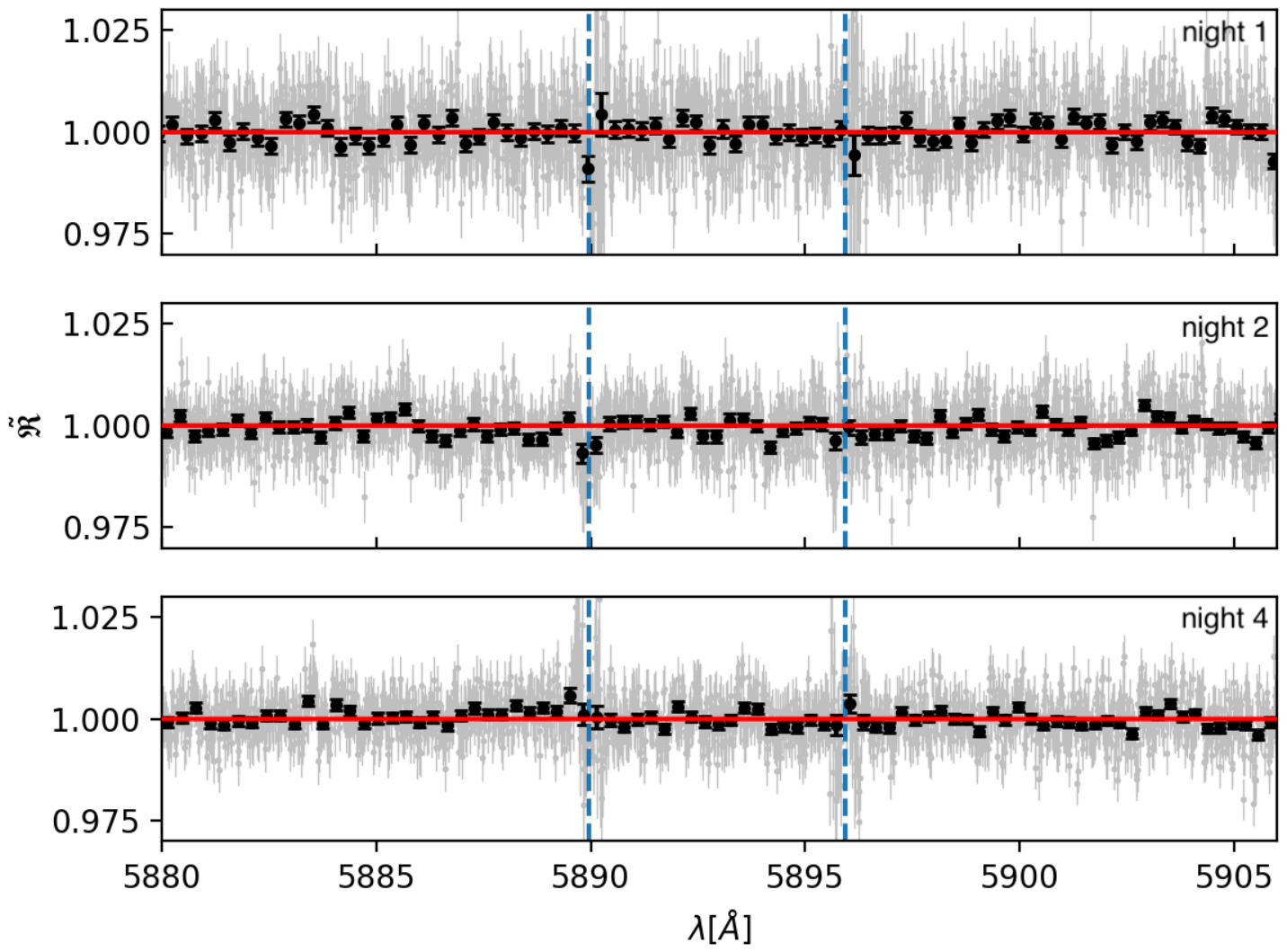}}
        \caption{HARPS transmission spectrum of WASP-127b for the sodium doublet in the planetary rest frame for all three nights separately. See Figure \ref{fig:transspectrum} for details.}
        \label{fig:transspectrumnights}
\end{figure*}

\begin{figure*}[htb]
\resizebox{\textwidth}{!}{\includegraphics[trim=0.0cm 6.5cm 0.0cm 6.5cm]{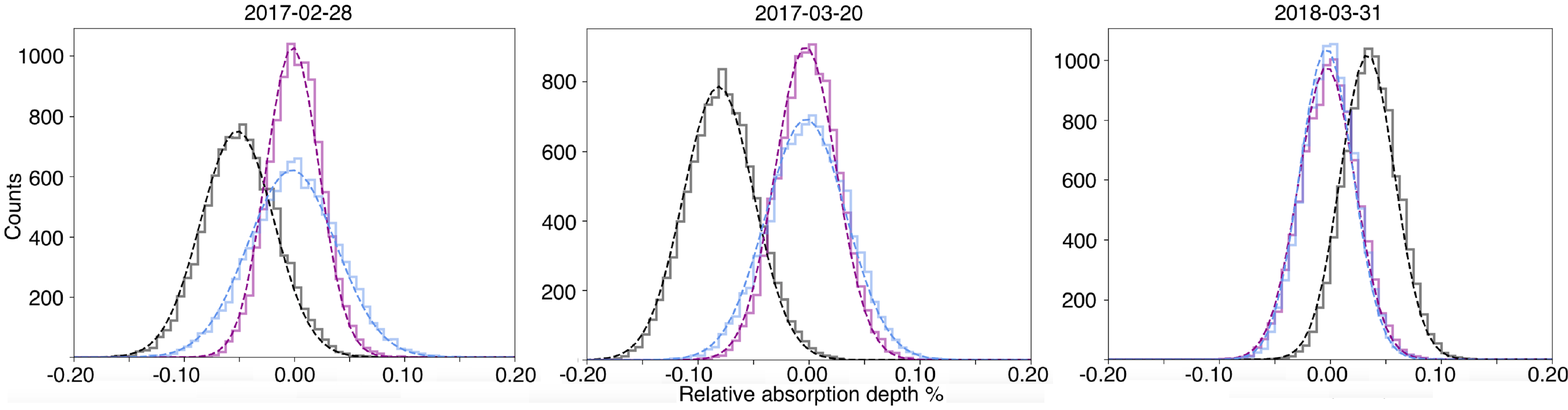}}
        \caption{Distributions of the bootstrapping analysis for the $12$ \r{A} passband for the three used transits (night one, two, four) for 20000 random selections. The `in-in' (magenta) and `out-out' (light blue) distributions are centred at zero (no planetary sodium detection) and the `in-out' distribution is shown in black.}
        \label{fig:random}
\end{figure*}

The resulting transmission spectrum for the three nights combined can be found in Figure \ref{fig:transspectrum}. Using the Gaussian fit of the D2 line to estimate the detection level \citep[see ][ for more information]{Hoeijmakers2020} results in a signal for the D2 line of $0.729\pm0.454\%$, corresponding to $1.6\sigma$. Although an absorption can be seen at the line centre of the D2 line, it was not detected in the D1 feature, and thus fitting Gaussians to both lines simultaneously was not possible. To calculate the significance of the overall sodium detection despite this, we compared the absorption in a $12\AA$ window around the position of the sodium doublet (dashed blue lines in Figure \ref{fig:transspectrum}), with the absorption in reference bands outside the sodium feature, as introduced in \cite{Wy15}. In the dataset presented here, the absorption in the sodium band is $0.456\pm0.198\%$, corresponding to a significance of $2.3\sigma$.

However, splitting the transmission spectrum into each night separately (see Figure \ref{fig:transspectrumnights}) highlights that this feature stems largely from night one, with some absorption at the D2 line position in night two, but no indication of a sodium feature in night four. To quantify this assessment, we performed a Monte Carlo bootstrapping analysis based on \cite{Re08}. If the sodium detection is indeed produced by the planetary atmosphere and is not a spurious occurrence, no random selection of the spectra for a random in-transit and out-of-transit group of spectra should be able to reproduce the results. We created three different scenarios to test this hypothesis. An 'in-in' and 'out-out' scenario, where the random in-transit and out-of-transit spectra were all taken only from the real in-transit spectra or were all taken only from the real out-of-transit spectra, respectively. In the control case, 'in-out', the spectra were still drawn randomly, but for the random in-transit spectra only from the real in-transit spectra and vice versa. 

The relative absorption depth distributions for the three analysed transits can be found in Figure \ref{fig:random}. Both 'in-in' and 'out-out' are centred at $0.0\%,$ as expected, meaning that any sodium detection cannot be reproduced solely from either the in-transit spectra or the out-of-transit spectra. However, the 'out-out' distribution has a large spread for night one (Figure \ref{fig:random}), which results in a false-positive likelihood for the sodium detection in night one of $2.5\%$ \citep{Re08}. The large spread of the 'in-out' sample in the same night highlights the high noise in this specific night, which allows for a wide range of possibilities. No detection could be confirmed for night four. 

\noindent Our results therefore show that a detection of sodium from this HARPS dataset can neither be confirmed nor confidently ruled out. It also remains unclear whether our inability to fit the D1 line stems only from the low S/N of the spectrum or if a physical mechanism reduces the strength of the D1 line disproportionally. Assuming that there is a sodium feature for WASP-127b, we propose a similar line of thought as for the ultra-hot Jupiter WASP-121b, a planet with a confirmed sodium detection and an interesting line ratio between the D2 and D1 line \citep{Hoeijmakers2020}: If sodium exists in the atmosphere of the planet and is extended enough, yet contained in the Roche radius, and additionally, if the number density of sodium is low, the outflow of sodium is optically thin in the line cores of the sodium doublet. This scenario is likely for WASP-127b, given that it is the puffiest exoplanet found to date. As shown in \cite{Draine2011}, the line depth ratio of the D1 and D2 line depends on the ratio of the oscillator strengths in the thin limit, which is approximately $2$ for the sodium doublet. \cite{Gebek2020} linked the theoretical findings from \cite{Draine2011} and the puffy nature of some exoplanets and showed that when optically thin sodium gas is assumed across both line cores, the line ratio of the sodium doublet approaches 2 \citep[see also Appendix A in][]{Hoeijmakers2020}. This hypothetical, optically thin, toroidal envelope of sodium around the planet could produce the line ratio needed to detect the D2 line, but might obscure the D1 line for WASP-127b.

\section{Comparison with previous work}

Placing our result in context with previous studies of WASP-127b highlights the need for further investigation of a potential sodium signature. WASP-127b has been observed multiple times at low resolution with ground- and space-based facilities. The first of these studies detected a strong Rayleigh slope and features compatible with TiO and VO in a most likely cloud-free atmosphere \citep{Palle2017}. However, sodium was detected tentatively in \cite{Palle2017} and was then claimed as a detection at the $5\,\sigma$ level in the GTC follow-up \cite{Chen2018}.  \cite{Chen2018} detected potassium and lithium in addition to sodium, and the sodium detection was used to constrain the sodium abundance to super-solar.
More recently, \cite{Spake2019} studied WASP-127b using the Hubble and Spitzer Space Telescopes and detected $\rm H_2O$, $\rm CO_2$, while additionally confirming the sodium detection from the first two studies with a super-solar abundance, despite an overall weaker signature in comparison, which might be due to the different resolutions of the instruments.

Lastly, as discussed in Section \ref{sec:stellar}, Z19 presented a sodium detection based on the first two nights of this dataset and claimed a confirmation of the results presented with GTC instruments \citep{Palle2017, Chen2018}.  
We re-sampled our analysis on the HST/STIS wavelength grid from Figure 21 in \cite{Spake2019} showing all previously taken low-resolution datasets in $R_p/R_\star$. In the bin containing the sodium doublet, our data correspond to $0.1033\pm0.0009 R_p/R_\star$, which is lower than the sodium detection in \cite{Spake2019} or \cite{Palle2017} and \cite{Chen2018} (see Figure 21 in \cite{Spake2019}), but they are compatible at the $1\,\sigma$ level.

\section{Conclusion}
\label{sec:conclusion}

We studied the puffy super-Neptune WASP-127b in photometry and spectroscopy. We analysed seven transits from EulerCam and four transits from TESS and redetermined the stellar parameters from spectroscopy. The obtained updated system parameters were subsequently used to create the transmission spectrum.
The spectroscopic data of four transits of WASP-127b were obtained with the HARPS spectrograph during the HEARTS survey, of which one transit was rejected because the out-of-transit baseline was insufficient.
 We showed that the previous analysis published in Z19 of two out of the four nights from this dataset resulted in a false-positive detection of neutral sodium based on contamination by stellar sodium and low S/N residuals. Our analysis of three nights of the same dataset showed an overall absorption in the $12\AA$ band around the position of the sodium doublet at the $0.456\pm0.198\%$ level, corresponding to $2.3\sigma$. This result is compatible with previous work conducted with space- and ground-based instruments, but it remains a non-detection. Curiously, we were able to fit a Gaussian to the D2 line, but not to the D1 line. This could be due to low S/N of the data or alternatively, if sodium exists in the atmosphere, due to a toroidal envelope of sodium in the optically thin limit. This curious finding shows the power of high-resolution spectroscopy in studying line shapes and ratios that would remain unnoticed at low resolution. We set the 3$\sigma$ limit for the amplitude of a sodium detection with high-resolution spectrographs at $0.594\%$ for future studies of the existence and potential ratio of the sodium lines. WASP-127b is a challenging target and a cautionary tale regarding the effect of low S/N remnants and telluric contamination. Any future analysis of this target has to carefully account for these complications.


\begin{acknowledgements}
This project has received funding from the European Research Council (ERC) under the European Union's Horizon 2020 research and innovation programme (project {\sc Four Aces}; grant agreement No. 724427).
This work has been carried out within the frame of the National Centre for Competence in Research `PlanetS' supported by the Swiss National Science Foundation (SNSF). FPE and CLO would like to acknowledge the SNSF for supporting research with HARPS and EULER through the SNSF grants nr. 140649, 152721, 166227 and 184618. This work was supported by FCT - Funda\c c\~ao para a Ci\^encia e a Tecnologia through national funds and by FEDER through COMPETE2020 - Programa Operacional Competitividade e Internacionaliza\c c\~ao by the following grants: UID/FIS/04434/2019; UIDB/04434/2020; UIDP/04434/2020; PTDC/FIS-AST/32113/2017 \& POCI-01-0145-FEDER-032113; PTDC/FIS-AST/28953/2017 \& POCI-01-0145-FEDER-028953. A.W. acknowledges the financial support of the SNSF by grant number P400P2\_186765. HMC acknowledges financial support from a UK Research and Innovation Future Leaders Fellowship. N. A.-D. acknowledges the support of FONDECYT project 3180063. A. G. acknowledges support from the French CNRS and from the Agence Nationale de la Recherche (ANR grant GIPSE ANR-14-CE33-0018). A. G. has been supported by a grant from Labex OSUG@2020 (Investissements d’avenir – ANR10 LABX56).

We thank L. A. dos Santos for his comments and insights during the preparation of this work.
\end{acknowledgements}
%

\bibliographystyle{aa} 
\bibliography{WASP127bSodium}

%

\begin{appendix}
\onecolumn
\section{Low S/N residuals}
\label{app:2Dmaps}

\begin{figure*}[htb!]
\resizebox{\textwidth}{!}{\includegraphics[trim=0.0cm 9.0cm 0.0cm 9.5cm]{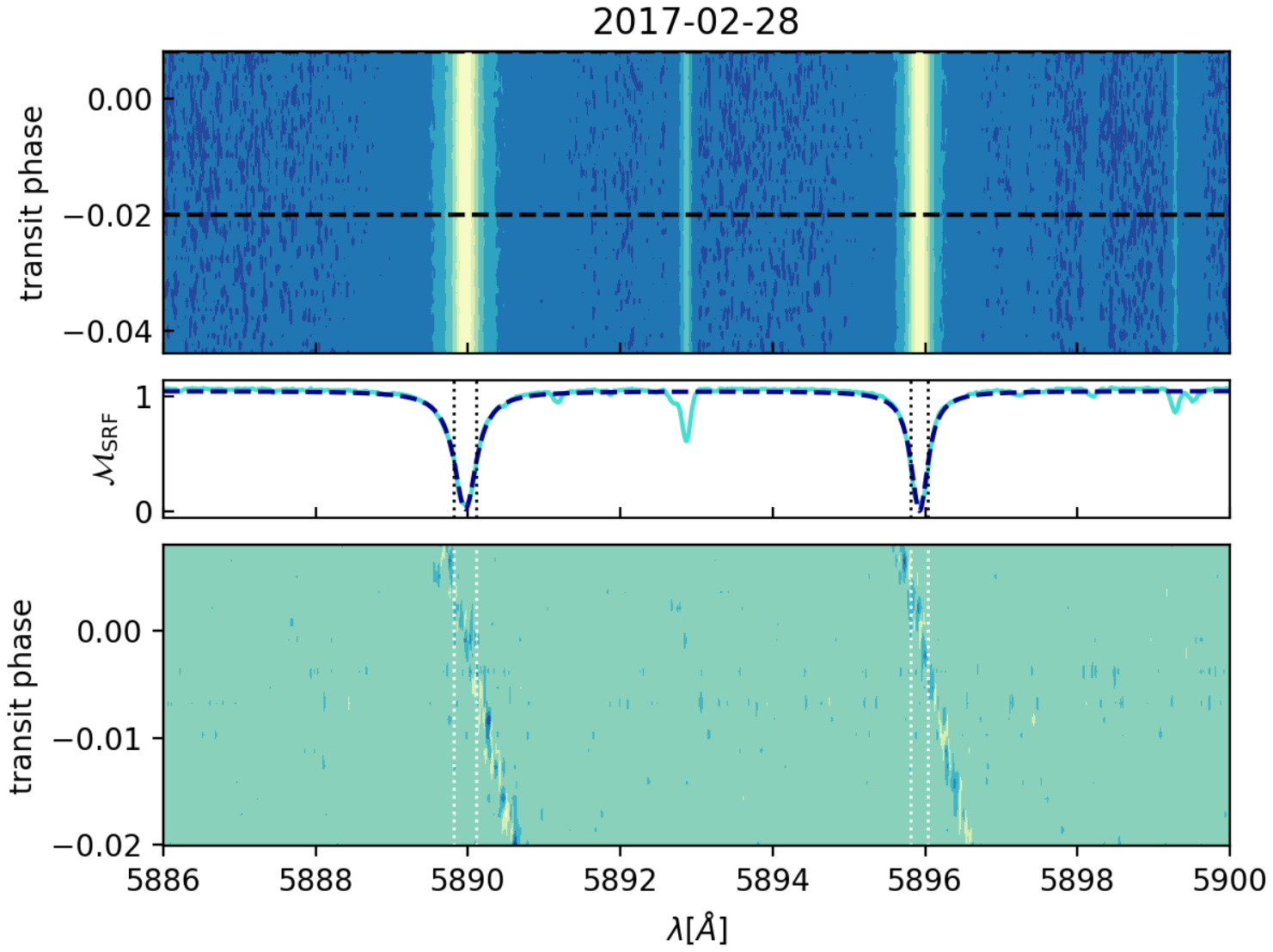}}
        \caption{Analysis of the overlap of the stellar and planetary sodium features for night one. The upper panel shows all spectra in the stellar rest frame (SRF) taken during the transit in the wavelength range of the sodium doublet stacked on top of each other to create a 2D map. The black dashed line indicate the ingress and egress (here only ingress for for this partial transit). The middle panel shows the sum over all spectra in the SRF, with a line profile fit as a dark blue dashed line. The FWHM is indicated as vertical, black dotted lines. The lower panel shows the spectra in the planetary rest frame (PRF), after division with the master-out spectrum. The position of the FWHM of the stellar sodium feature is indicated with dotted white lines. The low S/N residuals can be seen clearly as a strong trace in the lower panel.}
        \label{fig:2Dnight1}
\end{figure*}

\begin{figure*}[htb!]
\resizebox{\textwidth}{!}{\includegraphics[trim=0.0cm 9.0cm 0.0cm 9.0cm]{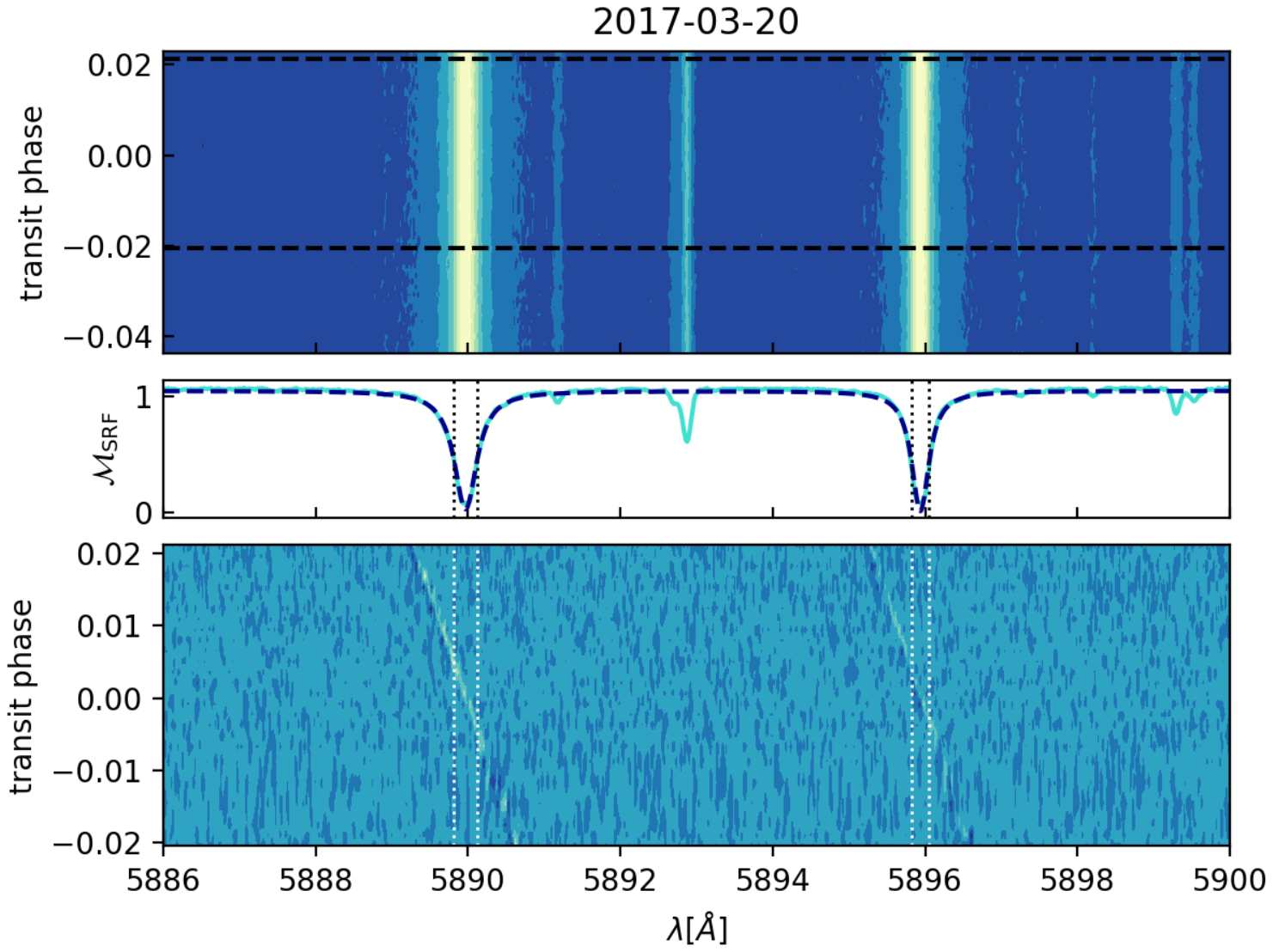}}
        \caption{Analysis of the overlap of the stellar and planetary sodium features for night two. See Figure \ref{fig:2Dnight1} for further details. The low S/N residual trace is visible for both the D2 and D1 line.}
        \label{fig:2Dnight2}
\end{figure*}


\begin{figure*}[htb!]
\resizebox{\textwidth}{!}{\includegraphics[trim=0.0cm 9.0cm 0.0cm 9.0cm]{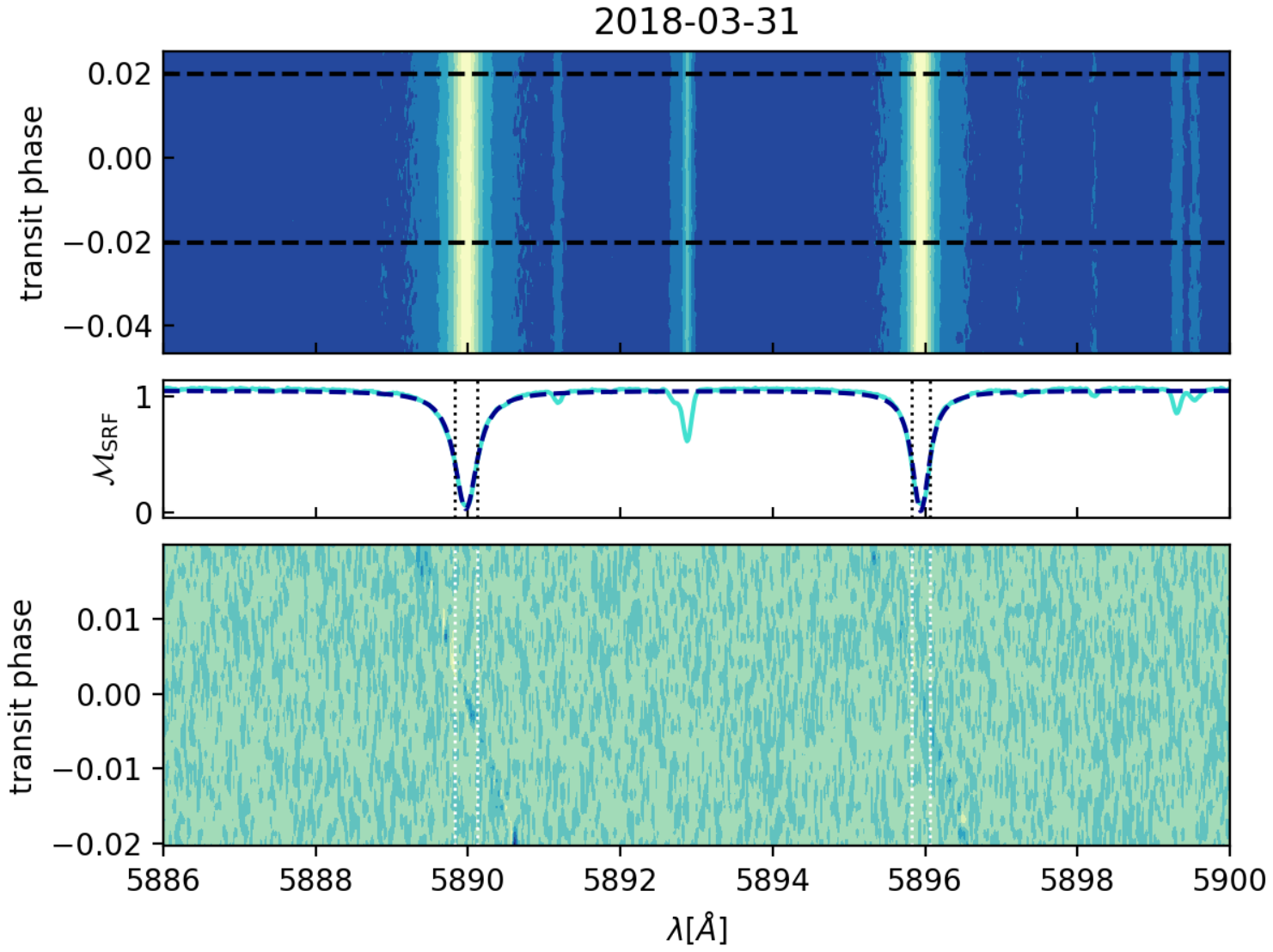}}
        \caption{Analysis of the overlap of the stellar and planetary sodium features for night four. See Figure \ref{fig:2Dnight1} for further details. The low S/N residual trace is weak for this night.}
        \label{fig:2Dnight4}
\end{figure*}

\end{appendix}

\end{document}